\documentclass{article}
\usepackage{arxiv2}
\usepackage{hyperref}
\usepackage{caption}
\usepackage{subcaption}
\usepackage{xcolor}
\usepackage{inputenc}
\usepackage{booktabs}
\usepackage{graphicx}
\usepackage{amsfonts}       
\usepackage{nicefrac}       
\usepackage{microtype}      
\usepackage{lipsum}
\usepackage{amsmath}
\usepackage{caption}

\usepackage{cleveref}
\definecolor{ggred3}{RGB}{248,118,109}
\definecolor{gggrn3}{RGB}{0,186,56}
\definecolor{ggblu3}{RGB}{0,191,196}
	
\title{Model-based recursive partitioning for discrete event times}

\author{
	Cynthia Huber\thanks{\texttt{cynthia.huber@med.uni-goettingen.de}} \\
	Department of Medical Statistics\\
	University Medical Center G\"ottingen\\
	G\"ottingen, Germany \\
	\And
	Matthias Schmid \\
	Institute of Medical Biometry, Informatics and Epidemiology\\
	Medical Faculty\\
	University of Bonn\\ 
	Bonn, Germany\\
	\And
	Tim Friede\\
	Department of Medical Statistics\\
	University Medical Center G\"ottingen\\
	G\"ottingen, Germany \\
}

\begin{document}
		\maketitle

		\begin{abstract}
			
			Model-based recursive partitioning (MOB) is a semi-parametric statistical approach allowing the identification of subgroups that can be combined with a broad range of outcome measures including continuous time-to-event outcomes. When time is measured on a discrete scale, methods and models need to account for this discreetness as otherwise subgroups might be spurious and effects biased. The test underlying the splitting criterion of MOB, the M-fluctuation test, assumes independent observations. However, for fitting discrete time-to-event models the data matrix has to be modified resulting in an augmented data matrix violating the independence assumption. 
			\\ 
			We propose MOB for \textbf{d}iscrete \textbf{S}urvival data (MOB-dS) which controls the type I error rate of the test used for data splitting and therefore the rate of identifying subgroups although none is present. MOB-ds uses a permutation approach accounting for dependencies in the augmented time-to-event data to obtain the distribution under the null hypothesis of no subgroups being present. 
			Through simulations we investigate the type I error rate of the new MOB-dS and the standard MOB for different patterns of survival curves and event rates. We find that the type I error rates of the test is well controlled for MOB-dS, but observe some considerable inflations of the error rate for MOB.
			To illustrate the proposed methods, MOB-dS is applied to data on unemployment duration.
		\end{abstract}


	
	\section{Introduction}

	Model-based recursive partitioning (MOB) \cite{mobZ, mobS} has become a popular approach for the identification of subgroups being similar with respect to the outcome of interest. The idea of the method is to improve the model fit by applying a tree-based algorithm and fitting separate models in each of the terminal nodes. MOB is applicable to a broad range of outcome types and is implemented in the R-package  \texttt{partykit} \cite{partykit}, making the use of the method flexible and user-friendly. Among others MOB is applicable to normally distributed and binary outcomes as node-wise fitting is based on the generalized linear modelling framework. The \texttt{partykit} \cite{partykit} package even provides the possibility of subgroup identification for continuous survival outcomes by adapting conditional inference trees 
	to censored data 
	\cite{mobS,Seibold_ALS, model4you}. \\
	The standard MOB approach treats the event time as a continuous random variable throughout. Although being appropriate in many settings, this strategy may not apply to situations where time is measured on a coarse scale (e.g.\@ when event times have been grouped or rounded). Ignoring the discreteness of the time-to-event data may lead to biased estimators and inaccurate predictions. Additionally, in discrete time-to-event data so called "ties", i.e.\@ observations having the same event time, can occur. However, methods for continuous event times commonly assume that ties are not present. \\
	Common examples of discrete event times include clinical studies with a fixed number of follow-up visits that do not allow for monitoring event times continuously over the whole follow-up period (thereby resulting in grouped survival data relating to a fixed number of intervals \cite{Schmid2016}) and studies involving {\em intrinsically discrete} duration times (such as the time to pregnancy measured by the number of menstrual cycles \cite{sundaram}). Monthly or annual surveys can also result in discrete measurements. Clearly, these types of data require specialized methods for discrete event times and should be modelled accordingly \cite{willettSinger, discreteSurv, schmidBergerWIRES}. Regarding tree-based modelling, Schmid et al \cite{Schmid2016} and Bou-Hamad et al \cite{Bou} proposed discrete-time recursive partitioning approaches with the purpose of prognostic subgroup identifaction. 
	Moradian et al \cite{rfDiscSurv} also introduced tree-based approaches and investigated random forest algorithms for discrete time-to-event data. Unlike MOB, these  methods are all based on impurity measures (such as the Gini criterion) to split the data but do not rely on formal statistical hypothesis testing to identify the subgroups.\\ 
	Generally, as the likelihood of a discrete time-to-event model is equivalent to the likelihood of a binary regression model \cite{willettSinger}, one may be tempted to apply MOB to discrete time-to-event data by using the standard implementation for binary outcomes. 
    In binary models with independent subjects, MOB controls the percentage of erroneously identified subgroups when in fact none is present corresponding to no instabilities in the parameters of the fitted binary model. This percentage is controlled by the test included in MOB's splitting criterion, namely the M-fluctuation test, and the use of multiplicity adjustments
	\cite{huber, sies,loh2019subgroup}. However, for fitting discrete survival models it is common to use an {\em augmented} data matrix. This means that subjects are represented by multiple rows
	in the data matrix (given their observed survival time exceeds one). Consequently, the assumption of independent observations made for MOB is violated and the asymptotic theory of the M-fluctuation test 
	is not valid in this case. Specifically, the type I error rate of the test underlying the splitting procedure is not  properly controlled when applying MOB to discrete time-to-event data.
	\\
	In this article we  demonstrate that ignoring dependencies in the augmented data matrix leads to a systematic inflation of the type I error rate when the standard MOB approach with binary outcome is applied to a set discrete time-to-event data.
	Furthermore, to address this problem, we propose an adjusted MOB algorithm (in the following referred to as {\em MOB-dS}) that is specifically tailored to modelling \textbf{d}iscrete \textbf{s}urvival data. Since the asymptotic theory for the M-fluctuation test used as MOB's splitting criterion is not valid for discrete time-to-event models, we use a permutation procedure
	for obtaining the distribution of the test statistic under the null hypothesis of no subgroups. For binary outcomes, a special case of discrete time-to-event data, the recent publication by Wolf et al \cite{VT} proposes a permutation approach to control the type I error rate in the Virtual Twin framework developed to detect heterogeneous treatment effects.
	\\
	The remainder of the article is organized as follows: In Section \ref{methods} we introduce the notation and present the proposed MOB-dS method. Section \ref{simulation} contains the results of a Monte Carlo simulation study to systematically evaluate the performance of the proposed procedure. In Section \ref{Application} we illustrate our proposal by applying it to data on unemployment duration \cite{Ecdat}. Section \ref{discussion} concludes the paper, providing a summary and discussion of the main findings.
	
	\section{Methods}\label{methods}
	
	\subsection{Notation}
	Throughout this article we will consider data consisting of $N$ subjects. The outcome is a discrete event time taking values in $\{ 1,\ldots,K \}$, which might result from $K$ underlying intervals $[a_0,a_1),[a_1,a_2),\ldots,[a_{K-2},a_{K-1}),[a_{K-1},\infty)$. In the latter case, an event time $T=t$ means that the event of interest occurred in the interval $[a_{t-1},a_t)$. 
	The observed time $\tilde{T}=\min(T, C)$, corresponding to right-censored data, is based on the true event time, denoted by $T\in \{ 1,\ldots,K \}$, and the true censoring time, denoted by $C\in \{ 1,\ldots,K \}$.
	The event time $T$ and censoring time $C$ are assumed to be independent (``random censoring''). Furthermore, the status indicator $\delta := I(T\leq C)$ indicates whether the observed survival time $\tilde{T}$ is right-censored ($\delta=0$) or not ($\delta=1$).\\
	We additionally consider two disjoint sets of covariates, namely  $\mathbf{X}=(X_1,\ldots,X_p)^T$ used by MOB to fit the regression models in the terminal nodes and $\mathbf{Z}=(Z_1,\ldots,Z_q)^T$ used for partitioning the data. The corresponding sets of covariate values for subject $i$ are denoted by $\mathbf{x}_i=(x_{i1},\ldots,x_{ip})^T$ and $\mathbf{z}_i=(z_{i1},\ldots,z_{iq})^T$, respectively.
	The available data consists of the independent vectors $\mathcal{D}=(\tilde{t}_i,\delta_i,\mathbf{x}_i^T,\mathbf{z}_i^T)$, $i=1,\ldots,N$.
	
	\subsection{Discrete time-to-event models and MOB}\label{basics}

	\subsubsection{Discrete time-to-event models}
	For discrete event times a wide range of parametric regression models is available~\cite{discreteSurv}. These models are usually based on the {\em discrete hazard function}, which has the form 
	$\lambda(t|\mathbf{X})=P(T=t|T\geq t,\textbf{x})$, describing the conditional probability of the event at time point $t$ given survival until $t$ \cite{discreteSurv}.
	The parametric discrete hazard model is defined by $g(\lambda(t|\mathbf{x}))=\gamma_{0t}+\boldsymbol{\mathbf{x}^T\boldsymbol{\beta}}$, where $g(\cdot)$ is a monotonic link function that relates conditional survival probabilities to the covariates $\mathbf{x}$ by a vector of regression coefficients $\boldsymbol{\beta}$. The set $\boldsymbol{\gamma_{0}}=\{\gamma_{01},\ldots, \gamma_{0,K-1}\}$ defines a covariate-free ``baseline'' trend (for $\mathbf{x}=0$). 
	Differences in discrete hazard models arise from the choice of the link function and assumptions concerning the baseline coefficients.
	Separate intercepts for each $t$ correspond to time-varying baseline hazards and are a common choice for discrete time-to-event models if $K$ is reasonably small.\\
	Defining $h = g^{-1}$, the parametric discrete hazard model becomes
	\begin{align}
		\lambda(t|\mathbf{x})= h(\gamma_{0t} + \mathbf{x}^T\boldsymbol{\beta} ),\nonumber\\ 
		g(\lambda)= \gamma_{0t}+ \mathbf{x}^T\boldsymbol{\beta} . \label{GLM}
	\end{align}	
	Common choices for $h(\cdot)$ are the logistic distribution function $h(\eta) = \exp(\eta)/(1 + \exp(\eta))$  or the Gompertz distribution $h(\eta) = 1 - \exp(- \exp(\eta))$. 
	The discrete time-to event models using the logistic or the Gompertz distribution are called proportional continuation ratio models or grouped proportional odds models, respectively. The grouped proportional odds model is a discretized version of the Cox proportional hazard model \cite{discreteSurv}.\\
	As the likelihood corresponding to Equation \eqref{GLM} is equivalent to the likelihood of a binomial model distinguishing whether the event took place at time $t$ or not (given $T\geq t$), the generalized linear modelling framework can be used for modelling discrete hazards $\lambda(t|\mathbf{x})$  \cite{Schmid2016, discreteSurv}.
	For uncensored and censored subjects, the response vectors of length $\tilde{T}_i$ that corresponds to the observed survival time of subject $i$ are $y_i=(0,\ldots,0,1)$ and $y_i=(0,\ldots,0)$, respectively.
	More specifically, the binary response $y_{it}$ of subject $i$ with $t = 1, \ldots, \tilde{T}_i$, is defined as 
	\begin{equation}
		y_{it}=\left\{
		\begin{array}{ll}
			1 \text{ if } t=\tilde{T}_i \text{ and } \delta_i=1,\\
			0 \text{ else, } 
		\end{array}
		\right.
		\label{binary}
	\end{equation}
	implying that the values $y_{it}$ are equal to $1$ if subject $i$ had an event in $[a_{t-1},a_t]$ and zero otherwise.
	
	Consequently, the design matrix for the binary regression model consists of $\tilde{T}_i$ rows for subject $i$, resulting in an {\em augmented data matrix} with $\sum_i \tilde{T}_i$ rows. Additionally the augmented data matrix contains a column of time indices for each subject $i$, given by $(t^*_1,\ldots , t^*_n)^\top$ with $t^*_i=(1,2,\ldots,\tilde{T}_i)$, $i=1,\ldots, n$. This column, when considered as a factor variable, is needed for estimating time-dependent intercepts as in Equation \eqref{GLM}. The covariate vectors $\mathbf{x}_i$ and $\mathbf{z}_i$ are duplicated subject-wise in each row of the augmented data matrix, as we do not assume time-varying covariates. The subject-wise duplicated vectors are denoted by $\mathbf{x}_i^*$ and $\mathbf{z}_i^*$, respectively.\\
	In R, the augmented data matrix 
	can be obtained using the function dataLong() of the add-on package \texttt{discSurv} \cite{discSurv_R}. Note that number of rows in the augmented data is denoted by $n$, which is (usually) considerably larger than the number of subjects $N$. The augented data matrix is defined by $\mathcal{D}_A= (y_k, t^*_k,\mathbf{x}_k^{*T},\mathbf{z}_k^{*T})$  for $k=1,\ldots, n$. 
	
	\subsubsection{MOB}
The idea underlying MOB is to improve the model fit by partitioning the data with respect to some covariates denoted $\mathbf{Z}$ and fitting separate local models of the form of Equation \eqref{GLM} within each of the resulting partitions. Consequently, these separate models have subgroup specific model parameters, e.g.\@ $\boldsymbol{\gamma}_j$ and $\boldsymbol{\beta}_j$ in subgroup $j$. As other tree algorithms MOB starts with all data, referring to the root node, and employs a splitting procedure to identify the splitting covariate and its cut-off value. The splitting procedure is defined by {\em M-fluctuation tests} \cite{mobZ,Zeileis2007}:  It tests whether the model parameters are stable across subgroups defined by splits in $\mathbf{Z}$, and the covariate associated with the highest parameter instability is selected as splitting variable. The test assesses systematic deviations of the score function, i.e. the gradient of the log-likelihood for Equation \ref{GLM}, evaluated at the estimated parameter from its mean being zero \cite{mobZ}.\\
A more detailed description of the M-fluctuation test for discrete event times can be found in Section \ref{MOBdS}.
The subgroups are further split by repeatedly applying the splitting procedure to the subgroups resulting from previous steps until a stopping criterion (e.g.\@ minimum required sample size in each subgroup) is met. The algorithm for MOB is shown in Figure \ref{fig:MOB_algo}, where Step 4 is highlighted inred. Step 4 of MOB (red) and MOB-dS (blue) described in Section \ref{MOBdS} are different.

\subsection{MOB-ds: Modification of MOB for the analyzing discrete time-to-event data}\label{MOBdS}

The use of binary regression for fitting discrete time-to-event models suggests that MOB can also be applied directly to discrete time-to-event data. However, as we will demonstrate below, using the standard M-fluctuation test to split the rows of the augmented data matrix may result in spuriously identified subgroups. This is because the M-fluctuation test assumes independent observations, which is usually not true for the rows of the augmented data matrix. To address this problem, we propose a permutation approach accounting for the structure of the augmented data. \\
For simplicity of presentation and without restriction of generality, the root node is considered in the following, but the results can analogously be applied to each node $j$. For readability, we omit therefore the index $j$. 
The splitting criterion is based on the model parameters ${\boldsymbol{\gamma}}$ and ${\boldsymbol{\beta}}$ of model \eqref{GLM} which are estimated by maximizing the log-likelihood $\Psi(\mathbf{y,x},\boldsymbol{\theta})$ with $\boldsymbol{\theta}=(\boldsymbol{\gamma,\beta})$. The score function corresponding to the log-likelihood is defined by $\psi(\mathbf{y,x},\boldsymbol{\theta})=\frac{\partial \Psi(\mathbf{y,x},\boldsymbol{\theta})}{\partial \theta}$. 
Instability over the considered variables $\mathbf{Z}$ is assessed by  M-fluctuation tests \cite{Zeileis2007, Zeileis2005}. The test assesses systematic deviations of the score function evaluated at the estimated parameter $\hat{\psi}=\psi(\mathbf{y,x},\hat{\boldsymbol{\theta}})$ from its mean being zero \cite{mobZ}.\\
For the $l$-th partitioning variable of $\mathbf{Z}$ the null hypothesis

	\begin{equation}\label{h0}
	    	H_0^{\theta,l}: \boldsymbol{\theta}_k=\boldsymbol{\theta}_0, \quad k=1,\ldots,n
	\end{equation}
is tested against the alternative that (at least one component of) $\boldsymbol{\theta}_k$ varies across $Z_l$  
with  $n$ denoting the number of rows of the augmented data matrix. The parameter vector $\boldsymbol{\theta}_k$ ($k=1,\ldots,n$) is a row-specific vector (based on $\mathcal{D}_A$) of regression coefficients.
The empirical fluctuation process $W_l(m)$ captures these deviations:
\begin{equation}
W_l(m)=\hat{J}^{-1/2}n^{-1/2}\sum_{r=1}^{\lfloor nm \rfloor}h'(\mathbf{x}^*_r\hat{\theta})\frac{y_r-\hat{\mu}_r}{\hat{\mu}_r(1-\hat{\mu}_r)}\mathbf{x}^*_r, \quad 0\leq m \leq 1
\label{efp}
\end{equation}
with $\mathbf{x}^*_r$ denoting row $r$ of the augmented data matrix $\mathcal{D}_A$ ordered by the values of $Z_l$.
Therefore, the empirical fluctuation process is the partial sum process of the scores 
ordered by the variable $Z_j$, scaled by the number of the rows $n$ and $\hat{J}$, an estimate of the covariance matrix $\text{Cov}(\psi(Y,X,\hat{\theta}))$. For more details on the M-fluctuation test see \cite{Zeileis2007}.\\
Under the null hypothesis of the standard M-fluctuation test (assuming independent observations) the empirical fluctuation process converges to a Brownian bridge denoted by $W^0$ as $n\to\infty$ \cite{mobZ,Zeileis2007}.
For using the empirical fluctuation process as test statistic a scalar function $\zeta(\cdot)$ is applied to the empirical fluctuation process, resulting in $\zeta(W_l(\cdot))$. Its corresponding limiting distribution is $\zeta(W^0(\cdot))$. 
For instance, the supLM statistics by Andrews \cite{andrews} is used as scalar function for numerical $Z$ in the MOB algorithm. Its limiting distribution is a squared, k-dimensional tied-down Bessel process. Using the methods of Hansen \cite{Hansen} the approximate asymptotic p-values can be computed. For categorical $Z$ the weighted sum of the squared $L_2$ norm of the increments of the empirical fluctuation process over the observations in one of the categories of $Z$ is used as scalar function $\zeta$. The resulting test statistic is $\chi^2$-distributed \cite{mobZ}. The use of another function $\zeta$ for categorical variables is needed as a total ordering of the observations $\mathbf{z}$ is not possible due to ties.
	
	\begin{figure}[h]
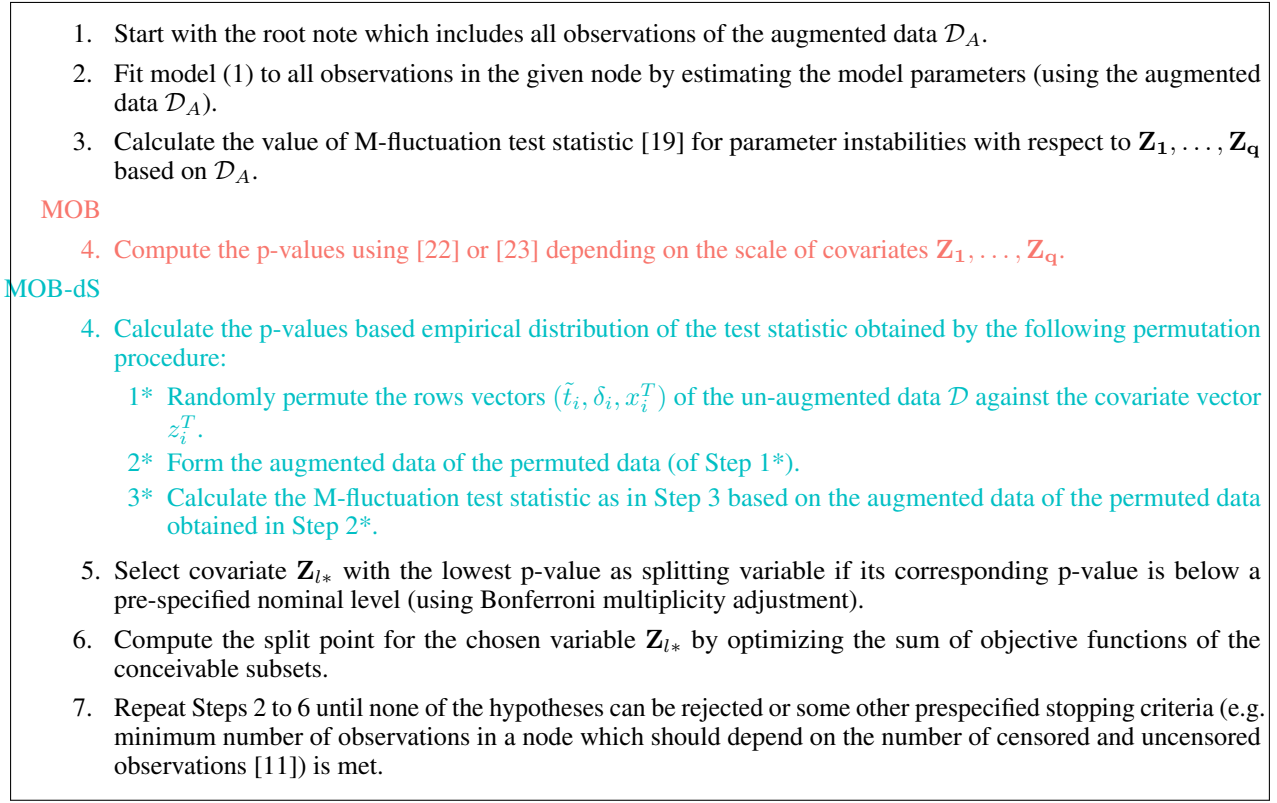

	    \centering
	    \noindent\fbox{\parbox{\textwidth}{
	\begin{enumerate}
		\item[1. ] Start with the root note which includes all observations of the augmented data $\mathcal{D}_A$.
		\item[2. ] Fit model \eqref{GLM} to all observations in the given node by estimating the model parameters (using the augmented data $\mathcal{D}_A$). 
		\item[3. ] 
    	Calculate the value of M-fluctuation test statistic \cite{Zeileis2007} for parameter instabilities with respect to $\mathbf{Z_1},\ldots, \mathbf{Z_q}$ based on $\mathcal{D}_A$.
    	\color{ggred3}
    	\item[MOB]
    	\item[4.] Compute the p-values using \cite{Hansen} or \cite{hjort} depending on the scale of covariates $\mathbf{Z_1},\ldots, \mathbf{Z_q}$.
    	\color{ggblu3}
    	\item[MOB-dS]
        \item[4.] Calculate the p-values based empirical distribution of the test statistic obtained by the following permutation procedure:
        \begin{itemize}
            \item[1*] Randomly permute the rows vectors $(\tilde{t}_i,\delta_i,x_i^T)$ of the  un-augmented data $\mathcal{D}$ against the covariate vector $z_i^T$.
            \item[2*] Form the augmented data of the permuted data (of Step 1*).
            \item[3*]Calculate the M-fluctuation test statistic as in Step 3 based on the augmented data of the permuted data obtained in Step 2*. 
        \end{itemize}
        \color{black}
		\item[5.]  Select covariate $\mathbf{Z}_{l*}$ with the lowest p-value as splitting variable if its corresponding p-value is below a pre-specified nominal level (using Bonferroni multiplicity adjustment). 
		\item[6. ] Compute the split point for the chosen variable $\mathbf{Z}_{l*}$ by optimizing the sum of objective functions of the conceivable subsets.
		\item[7. ] Repeat Steps  2 to 6 until none of the hypotheses can be rejected or some other prespecified stopping criteria (e.g.\@ minimum number of observations in a node which should depend on the number of censored and uncensored observations \cite{Bou}) is met.
	\end{enumerate}
	}}
	    \caption{Algorithm of MOB and MOB-dS. The two algorithms only differ in Step 4, which is highlighted in \color{ggred3} red for MOB \color{black} and in \color{ggblu3} blue for MOB-dS\color{black}.}
	    \label{fig:MOB_algo}
	\end{figure}

The algorithm of MOB-dS is described in Figure \ref{fig:MOB_algo}. MOB-dS differs in Step 4 from MOB: Whereas MOB computes the approximate asymptotic p-values 
using  methods  depending on the scale of the covariate \cite{Hansen, hjort}, MOB-dS uses a permutation approach for computing the distribution of the test statistic accounting for the augmented structure of the data used for discrete time-to-event data.
The asymptotic theory of the M-fluctuation test used in MOB is not valid for discrete time-to-event data because of the violation of the common assumption regarding independent observations \cite{Zeileis2007,mobZ} due to the use of the augmented data matrix.\\
The permutation approach in MOB-dS permutes the un-augmented data $\mathcal{D}$ mimicking the "null data" (Step 1*) before recalculating the test statistic on the augmented "null data" (Step 2* and Step 3*). 
A permuted sample of the un-augmented data $(\tilde{t}_i,\delta_i,\mathbf{x}^T_i,\mathbf{z}^T_i)$ is obtained by randomly permuting the row vectors  of $(\tilde{t}_i,\delta_i,\mathbf{x}^T_i)$ against the covariate vector $\mathbf{z}^T_i$ as this retains the overall effects of $\mathbf{X}$ and the correlations among $\mathbf{Z}$ while removing any marginal effects of the covariates $\mathbf{Z}$ on the outcome. The test statistic is then calculated on the permuted augmented data. The resulting distribution of the test statistics is used for calculating the p-value of the test statistics calculated on the original sample.

\section{Simulation study}\label{simulation}
To investigate the performance of MOB-dS, we carried out a simulation study structured by
the Clinical Scenario Evaluation (CSE) framework \cite{Benda2010}. Using the CSE framework, this section is divided into three parts, namely assumptions, options and metrics. 
Section \ref{assumption.sim} referring to CSE's assumptions defines the data generating model for the simulation. Section \ref{options.sim} for options describes the methods applied to the data and Section \ref{metrics.sim} for CSE's metrics specifies the criterion for evaluating the methods.\\
The aims of the study were to compare the type I error rate of the tests underlying the splitting criterion of both MOB and MOB-dS. MOB tests for parameter instabilities and therefore indirectly checks for the presence of subgroups as we assume that the parameters vary across subgroups if these are present.
		
\subsection{Assumptions: Simulation setting}\label{assumption.sim}

The data generated include a discrete survival outcome $\tilde{T}$, a censoring indicator $\delta$ and 13 covariates $\mathbf{Z}$ for each subject $i = 1, \ldots,N$. The 13 covariates $Z_1$,\ldots,$Z_{13}$ are drawn from a standard normal distribution. All covariates are correlated with either $\rho=0.1$ or $\rho=0.5$ (results in supplement). \\
For comparing the type I error of the test underlying the splitting criterion of MOB and MOB-dS, the time-to-event $T$ was generated using Equation \eqref{S0} 

\begin{align}
	g(\lambda)= \gamma_{0t}, \quad t=1,\ldots,K-1.
	\label{S0}
\end{align}

We considered the logit link as true link function $g(\cdot)$. The values of $\gamma_{0t}$ are constant across the covariate values of $\mathbf{Z}$ corresponding to data generated under $H_0$. For simplicity we did not consider additional variables $\mathbf{X}$ in the data generating model. 
The intercepts $\gamma_{0t}$  were varied across simulation settings. Censoring times were generated from a continuous exponential distribution and independent of the event time. The rates of the exponential distribution were chosen to achieve approximately 0\%, 20\% and 50\% censoring.  The sample size was set to  $100\cdot (K-1)$. 

\begin{figure}
	\centering
	\begin{subfigure}[b]{0.3\textwidth}
		\centering
		\includegraphics[width=1\linewidth]{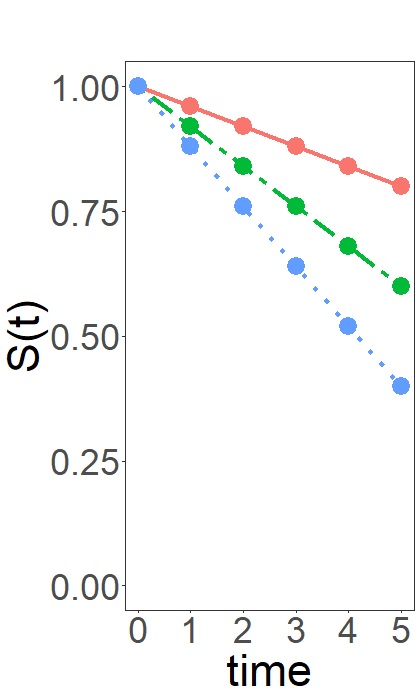}
		\caption{Setting A}
		\label{fig:S1A}
	\end{subfigure}
	\begin{subfigure}[b]{0.3\textwidth}
		\centering
		\includegraphics[width=1\linewidth]{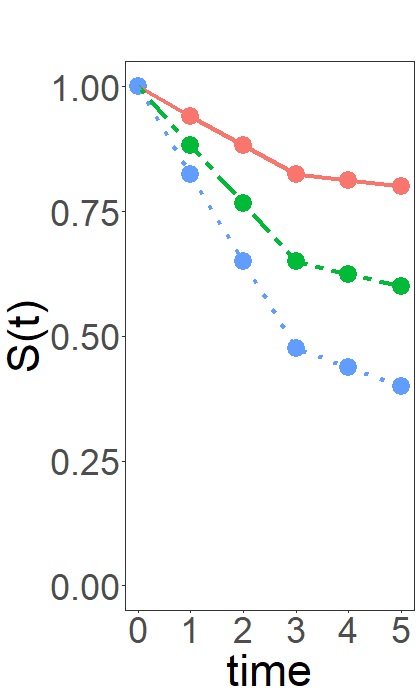}
		\caption{Setting B}
		\label{fig:S1B}
		\end{subfigure}
	\begin{subfigure}[b]{0.3\textwidth}
		\centering
		\includegraphics[width=1\linewidth]{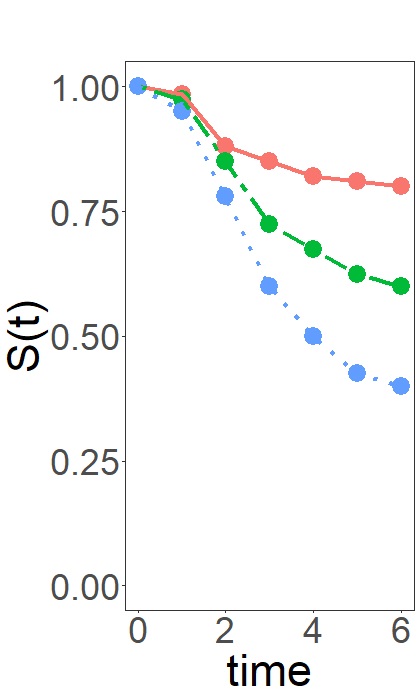}
		\caption{Setting C}
		\label{fig:S1C}
	\end{subfigure}
	\caption{Illustration of the true survival functions for settings with $K=7$. The different colours and line types correspond to settings with 80\% (solid), 60\% (dashed) and 40\% (dotted) events.}
	\label{fig:true_surv}
\end{figure}

For the choice of the values for $\gamma_{0}$ we considered different forms of survival functions with three different event rates namely 80\%, 60\% and 40\%. As an example, the different shapes of the survival function are shown for $K=6$ in Figure \ref{fig:true_surv}. Setting A considers a linear shape of the survival function, in setting B the events occur earlier and in setting C the events occur mainly in the middle of the observational period.
The baseline coefficients differ with differing shape of the survival function, differing discrete time points $K$ and considered event rates. We additionally varied the number of discrete time points $K \in \{4,5,\ldots, 11\}$.
\\
For each setting, 2000 Monte Carlo replications were generated.

\subsection{Options: Methods and their tuning parameters} \label{options.sim}

For the computation we used the software environment R \cite{rcore} and Version 1.2-1 of the R package \texttt{partykit}\cite{partykit} for applying MOB with discrete hazard models. All simulations were conducted using R 4.1.0 (64 bit) 
on the GWDG (Gesellschaft fuer wissenschftliche Datenverarbeitung mbH Goettingen) High Performance Cluster, located in Goettingen, Germany. 
For both MOB and MOB-dS, we considered a proportional continuation ratio model (PCRM) with time-dependent baseline coefficients as underlying model:  
	\begin{equation}
		\lambda(t)=\frac{\exp (\gamma_{0t})}{1+\exp (\gamma_{0t})}.
		\label{pcr_ext}
	\end{equation}
We used the 5\% level of significance for the test in the splitting criterion and the Bonferroni correction for multiplicity adjustment. As covariance matrix estimator in the parameter instability tests (see Equation \eqref{efp}) we used the sandwich estimator as the default covariance matrix estimator in the \texttt{partykit} package, the  outer product of gradients, can lead to numerical problems in the Cholesky decomposition especially in presence of sparse events.
We compared the type I error rate based on the asymptotic distribution of the M-fluctuation test used in MOB and the exact distribution obtained by permutation accounting for dependencies in discrete time-to-event data as used in MOB-dS. As we generated 2000 data sets for each setting, we used 1000 repetitions of the permutation approach within MOB-dS.\\

\subsection{Metrics: Performance criteria}\label{metrics.sim}
We compare the two methods by their simulated type I error rates of their splitting criteria. Additionally, we illustrate the right-hand tail probabilities from the distribution of the instability tests of the two methods. 

\subsection{Results} \label{results}

 \begin{figure}[h]
 	\centering
 	\includegraphics[width=1\linewidth]{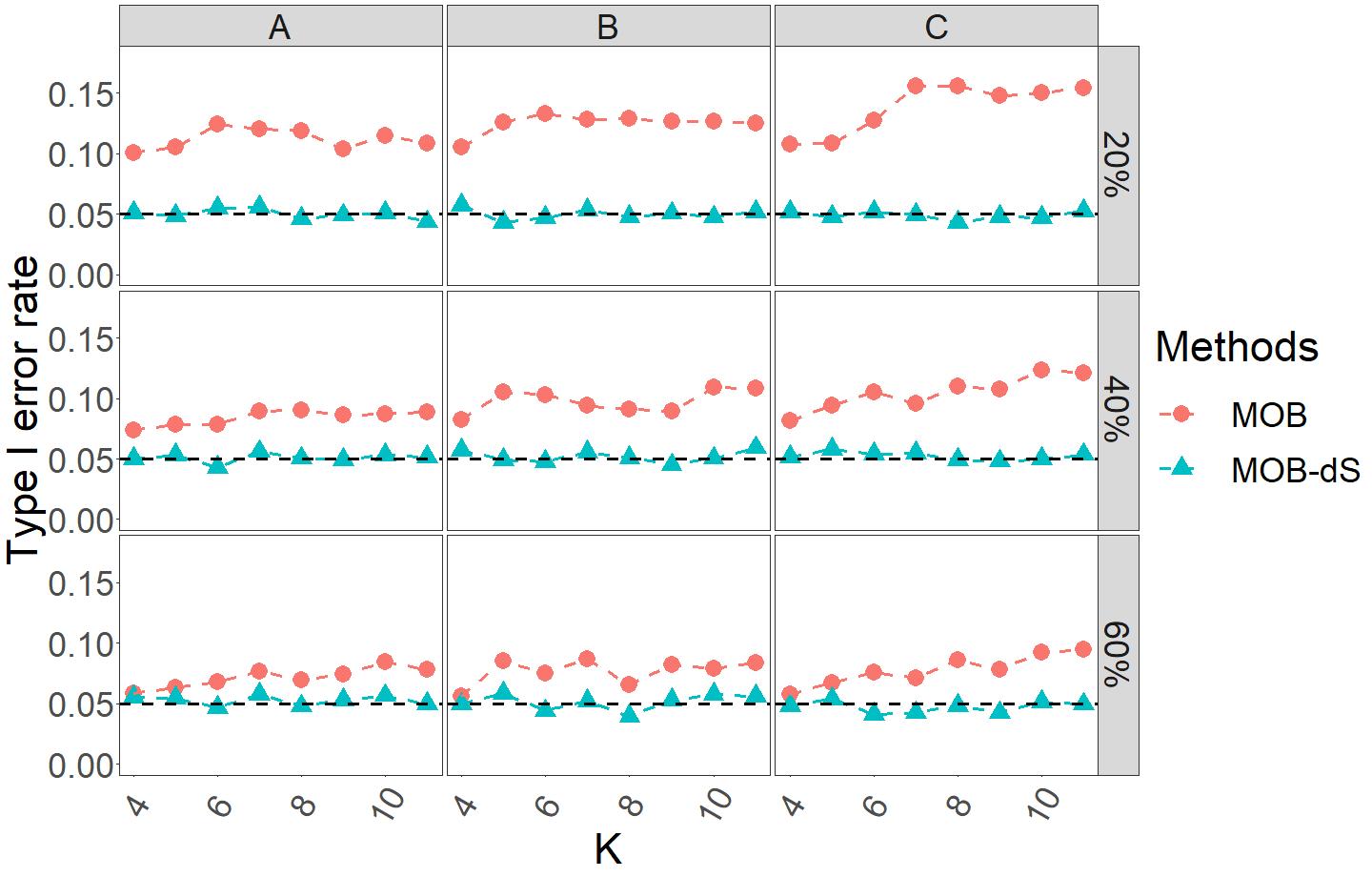}
 	\caption{Type I error rate for MOB and MOB-dS based on 2000 simulated data sets per setting with 20\% censoring and 1000 MC samples for MOB-ds. The rows represent settings with the three different expected event rates and the columns represent the different settings A, B and C as illustrated in Figure \ref{fig:true_surv}. The Monte Carlo simulation error is $\approx 0.4\%$. 
 	    }
 	\label{fig:type1errorCens02}
 \end{figure}
 
\begin{figure}[h]
 	\centering
 	\includegraphics[width=1\linewidth]{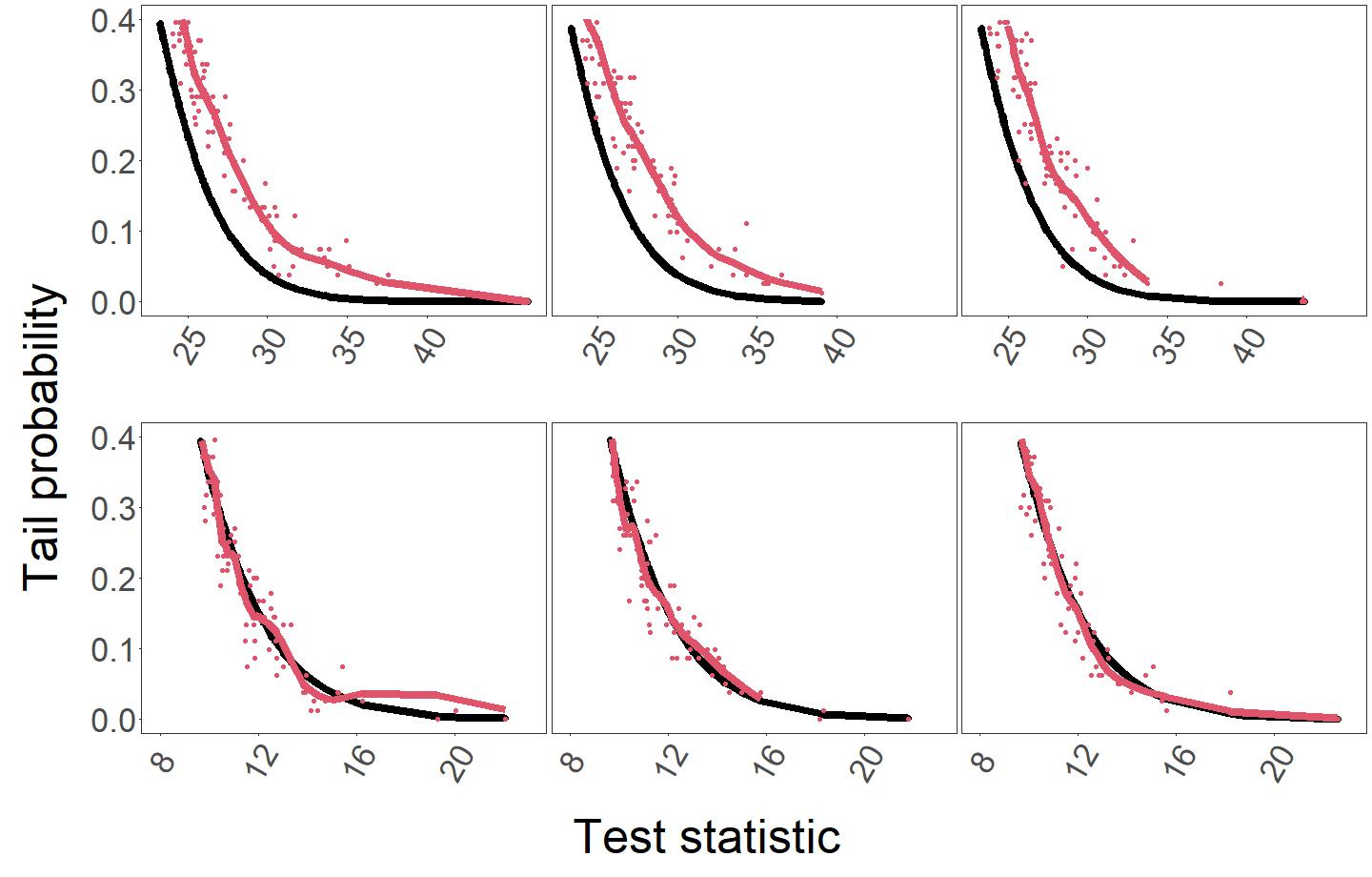}
 	\caption{ Right-hand tail probabilities from approximate asymptotic and sampling distributions of instability test in MOB (solid black) and MOB-dS (dashed red). The solid red lines correspond to the smoothed sampling distributions obtained by the LOESS estimator. Test statistics of three out of thirteen covariates are shown only. Distributions are shown for setting C with $K=8$ and 20\% events.
 	The sampling distribution used in MOB-dS is based on 1000 permuation samples. 
 	    }
 	\label{fig:distribution}
 \end{figure}

 Figure \ref{fig:type1errorCens02} shows the simulated type I error rate for different scenarios defined by the considered event probabilities, number of discrete time points $K$ and the different shapes of survival functions.
 The results of Figure \ref{fig:type1errorCens02} are based on scenarios with 20\% censoring. Each dot represents the simulated type I error rate of a single scenario based on 2000 Monte Carlo replications. Therefore, the Monte Carlo simulation error is $\sqrt{\frac{0.05*0.95}{2000})}\approx 0.4\%$.  Figure \ref{fig:type1errorCens02} consists of 9 single frames. The frames of the different rows show the scenarios with different event probabilities, whereas the columns correspond to settings with different shapes of the survival function. The type I error rate is plotted against the different discrete time points $K$.
It is seen that the distribution for the test used in MOB-dS (with 1000 MC samples) controlled the type I error rate well in all considered settings. In contrast, the type I error rate of MOB was inflated in most of the settings. Therefore, the probability of identifying non-existing subgroups using MOB is higher compared to MOB-dS. In settings with more events (third row of Figure \ref{fig:type1errorCens02}) and fewer discrete time points, the type I error rate of MOB is close to 5\%, e.g. for settings with $K=4$ and 60\% events the mean simulated type I error rate for MOB across the settings A,B and C is 5.75\%. Due to the higher event rate which do not only occur at the end of the observational period and fewer time points less dependencies in the augmented data matrix are present. This might contribute to the better type I error results of MOB in these settings.
The highest type I error rates for MOB are observable for settings with a large number of discrete time points and few events. For example, for a 20\% event rate the mean of the type I error rate across the settings A, B and C for no censoring was 12.7\%, whereas the simulated type I rate of MOB-dS was 4.9\%. The type I error rates for settings without censoring did not strongly differ from those with higher censoring probabilities (20\% and 50\%).\\
For a setting with higher type I error rates as for setting C (few events at the beginning and end of the observational period) with $K=8$ and 20\% events, we illustrate the right-hand tail probabilities from the distributions of the test statistics used in MOB (black) and MOB-dS (red), see Figure \ref{fig:distribution}. The first row corresponds to the tail probabilities for  the logistic discrete hazards model (\eqref{GLM} with $g(\lambda)=\log(\lambda / (1- \lambda))$ ). The second row shows the tail probabilities for test statistics based on logistic models using the censoring indicator $\delta$ as outcome ignoring the time to event information given in the simulated data. 
The approximate distribution used in MOB is due to Hansen \cite{Hansen}, whereas MOB-dS samples the distribution of the test statistic. For binary outcomes (second row) the approximation due to Hansen and the sampling distribution are very close. This is also notable in the type I error rates obtained by the Hansen approximation for the p-values of the instability test: For binary outcomes with independent observations the use of the Hansen approximation does not result in inflated type I error rates. However, the distribution of Hansen seems to be shifted compared to the sampling distribution for discrete hazards models, leading to the test used in MOB becoming anti-conservative. This confirms the results of the inflated type I error of the previous simulations. For discrete hazard models which use an augmented data matrix for fitting a logistic model the Hansen approximation seems therefore not suitable for approximating the asymptotic distribution of the instability test statistics.  

\section{Application}\label{Application}

To illustrate the use of MOB-dS, we apply the method along with MOB to data from the R package \texttt{Ecdat} on Unemployment Duration \cite{Ecdat}. 
\begin{table}[]
    \centering
    \begin{tabular}{c|c}
        & Overall (N= 3343) \\
        \hline
        Age in years & 34.00 (27.00;43.00)\\
        UI claim & 1848 (55.3\%)\\
        Eligible replacement rate  & 0.50 (0.40;0.52)\\
        Eligible disregard rate & 0.10 (0.05;0.15)\\
        Logarithmized weekly earnings in lost job& 5.68 (5.30;6.05)\\
        Tenure in lost job in years & 2.00 (0.00;5.00) \\
    \end{tabular}
    \caption{Characteristics of the subjects enrolled in the Unemplyment Duration data \cite{Ecdat} available in the R package \texttt{Ecdat}. Values are given as n (\%) or median (interquartile range) }
    \label{tab:application}
\end{table}

\begin{figure}[h]
 	\centering
 	\includegraphics[width=1\linewidth]{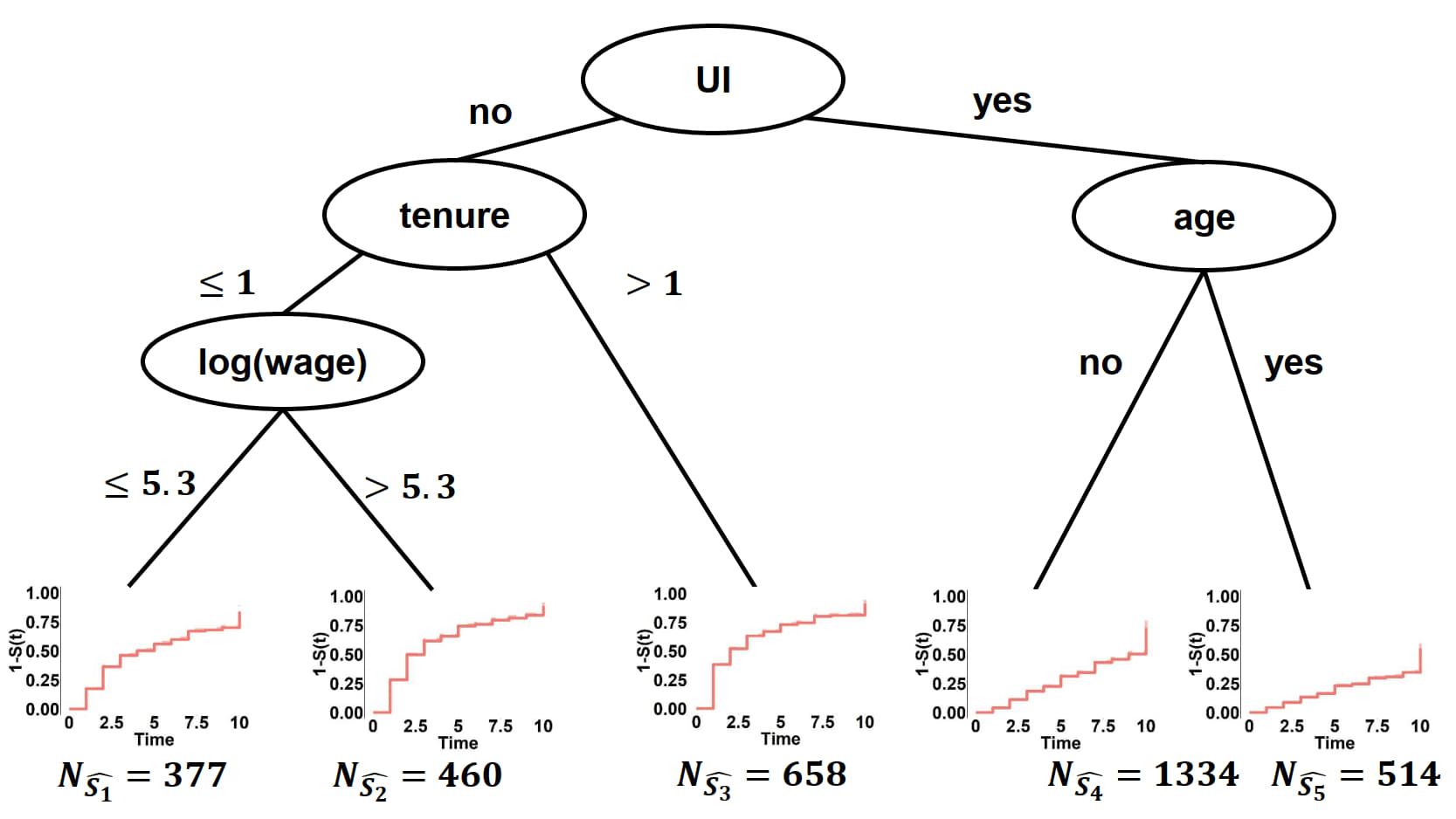}
 	\caption{Life table (dashed) and estimates of a proportional continuation ratio model with separate baseline hazards for each time point (solid) in subgroups identiﬁed with MOB-dS. There are hardly any differences between the two estimates. The covariates "UI", unemployment insurance claim, years "tenure" in lost job, "log(wage)", log transformed weekly wage and age were selected splitting variables by the MOB-dS approach.}
 	\label{fig:MOB_dS_Anwendung}
 \end{figure}

\begin{figure}[t]
 	\centering
 	\includegraphics[width=1\linewidth]{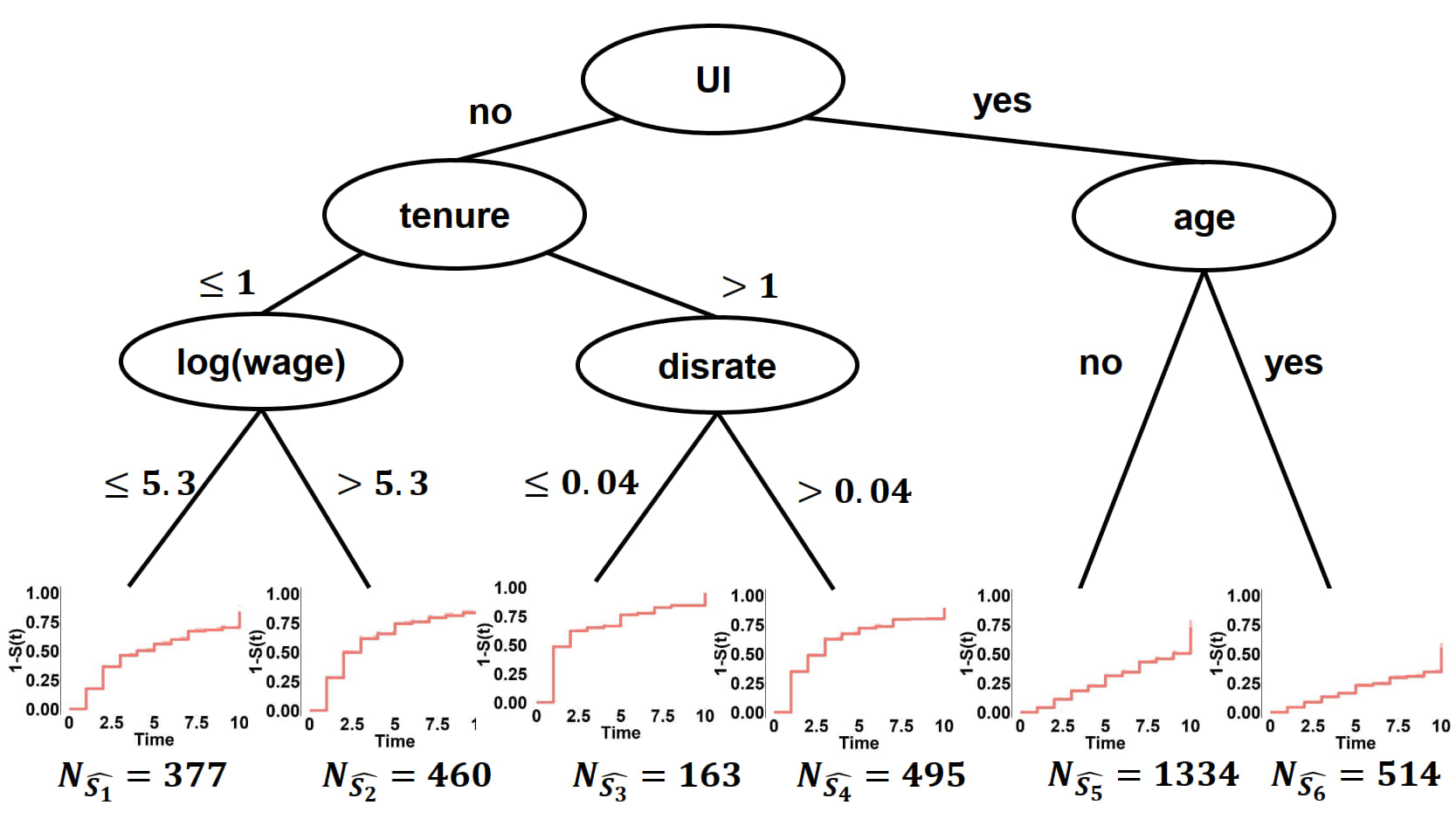}
 	\caption{Life table (dashed) and estimates of a proportional continuation ratio model with separate baseline hazards for each time point (solid) in subgroups identiﬁed with MOB. There are hardly any differences between the two estimates. The covariates "UI", unemployment insurance claim, years "tenure" in lost job, "log(wage)", log transformed weekly wage, "disrate", the eligible disregard rate, and "age" were selected splitting variables by the MOB approach.}
 	\label{fig:MOB_Anwendung}
 \end{figure}

The data set includes 3343 unemployed people and six covariates, namely age, unemployment insurance (UI) claim, eligible replacement rate, eligible disregard rate, weekly earnings in lost job (log transformed) and tenure in lost job. These characteristics are summarized in Table \ref{tab:application}. The time to re-employment, the event of interest, is measured on a discrete scale, namely two week intervals, ranging from 2 to 56 weeks. Due to few events in the time interval $[20,56)$, we collapsed the corresponding discrete times to one category. The discrete event time $t=1,\ldots,10$ is therefore based on the following intervals $[2,4),[4,6),\ldots,[18,20), [20,56)$. 
Figure \ref{fig:MOB_dS_Anwendung} illustrates the tree identiﬁed by MOB-dS with the underlying logistic discrete hazards model (Equation \eqref{GLM} with the logistic link function) based on 10000 Monte Carlo samples. 
Figure \ref{fig:MOB_Anwendung} shows the corresponding results for MOB. For both tree methods the maximum tree depth was set to four. The terminal nodes include the estimated cumulative probability for the favourable event ($1 - S(t)$), namely re-employment.
The life table estimates are illustrated with dashed lines whereas the solid lines represent the estimates obtained by the logistic discrete hazards model with separate hazards for each time point as used in both MOB and MOB-dS.  
MOB-dS identifies five whereas MOB identifies six subgroups with MOB-dS beeing a sub-tree of MOB. Therefore, four of the identified subgroups are identical for both methods. The difference arises from the splitting on covariate "disrate", the eligible disregard rate, which is identified by MOB but not by MOB-dS. 
The p-value for the variable "disrate" is close to 5\% for MOB and MOB-dS using 10000 resamples. The p-value for the split on "disrate" in the subgroup of subjects without unemployment insurance and more than one year tenure in their previous job is 4.7\% for MOB and 6.3 \% in MOB-dS.   \\
Both results suggest that people without an UI claim and less than a year tenure in the lost job are earlier reemployed if they have a higher wage. Subjects without UI claim and with more than one year tenure at the lost job also tend to be reemployed earlier than subjects with less than one year tenure at the lost job  
In MOB-dS people with an UI claim are additionally classified by the variable age. People older than 43 tend to find a new job later than people of 43 years or younger.
 
 We investigated MOB-dS with different size of MC samples ranging from 500 to 10000.
 The results of MOB-dS were stable across the different numbers of MC samples. MOB-dS does not split the data on "disrate" with any of the considered MC samples (500,1000,2000,3000,5000) and all the p-values for the split on "disrate" as in the result of MOB range from 0.06 to 0.09. 

\section{Discussion}\label{discussion}

MOB is a semi-parametric statistical approach for the identiﬁcation of subgroups that can be combined with a broad range of outcome types. Although fitting discrete time-to event models is feasible within the GLM framework, MOB is not readily applicable to discrete time-to-event data. We showed that the M-fluctuation test with the Hansen approximation which is commonly used as splitting criterion in MOB with binary outcome, is not suitable for discrete time to event data. In our simulations we demonstrated that the type I error rate of this test used as MOB's splitting criterion exceeded the 5\% significance level in most of the considered settings, leading to an inflated type I error rate and to the identification of spurious subgroups. This is due to the assumption of independent observations made for the M-fluctuation test, which is violated by the use of the augmented data matrix for discrete time to event modelling neither the asymptotic distribution nor the exact finite-sample permutation distribution obtained by permuting the augmented data matrix  (results not shown) results in a controlled type I error rate. The MOB-dS method proposed in this work, is an adaption of MOB tailored to the application to discrete time-to event data. The p-values of MOB-dS are obtained by using the permutation distribution using Monte Carlo samples of the un-augmented data. The simulations showed that MOB-dS controls the type I error rate better than MOB.\\
Although the type I error rate is controlled by MOB-dS it is computationally expensive as a sufficiently large number of permutations is needed to approximate the distribution of the test statistic. One approach to reduce the computation time would be to change the test statistics and to use asymptotic distributions as in MOB for the application with discrete time to event outcomes. Moreover, it is also of interest to investigate how well the subgroups are identified by MOB-dS which is subject of future research.


\appendix
\section{Additional figures}
\begin{figure}[p]
 	\centering
 	\includegraphics[width=1\linewidth]{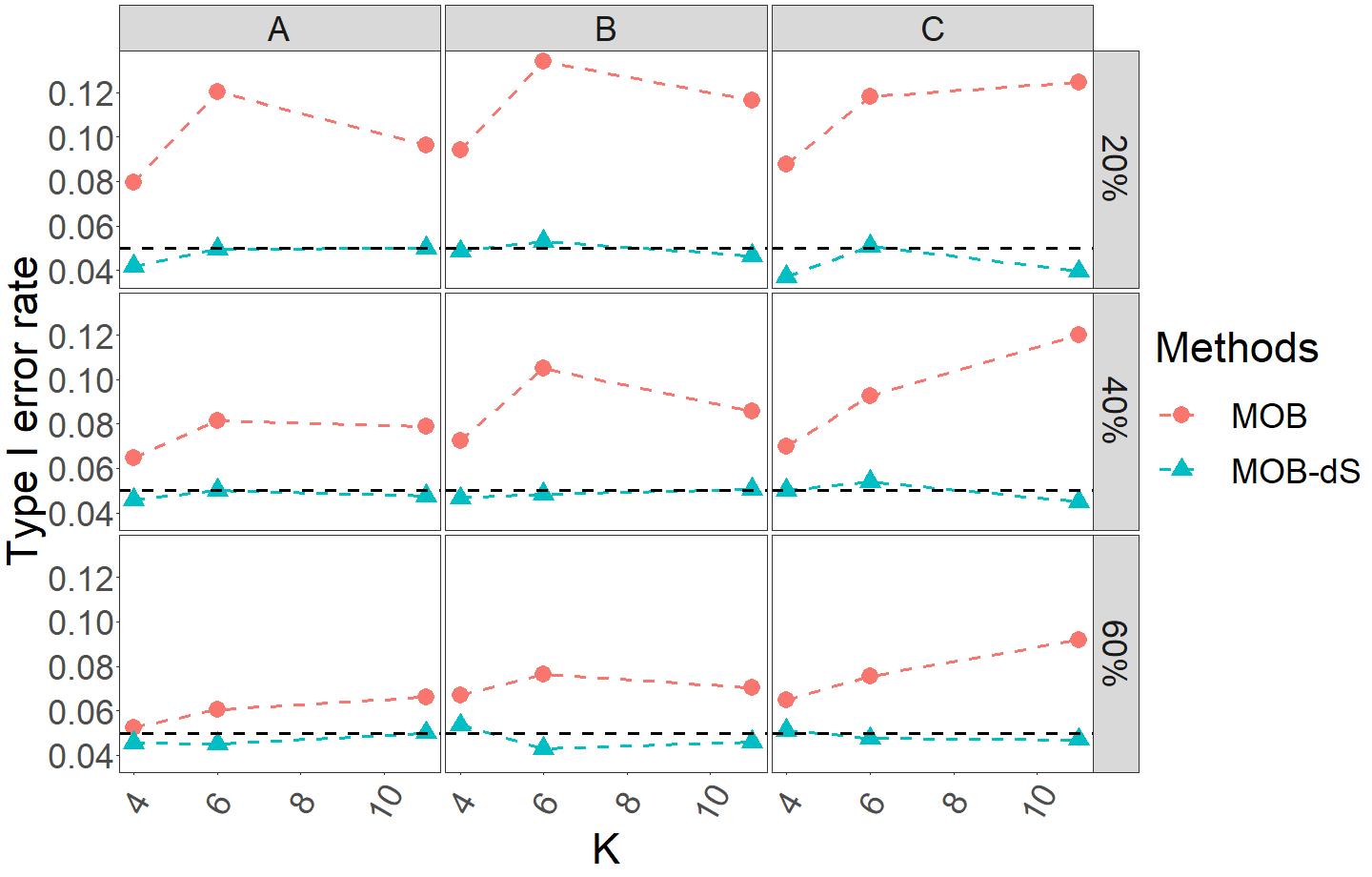}
 	\caption{Type I error rate for MOB and MOB-dS based on 2000 simulated data sets per setting with 20\% censoring and 1000 MC samples for MOB-ds. The covariates $\mathbf{Z}$ are correlated with $\rho=0.5$. The rows represent settings with the three different expected event rates and the columns represent the different settings A, B and C as illustrated in Figure \ref{fig:true_surv}. The Monte Carlo simulation error is $\approx 3\%$. }
 	\label{fig:TypeIerror_rho05}
 \end{figure}

For simulation scenarios in which the correlation of the variables $\mathbf{Z}$ is set to $\rho=0.5$ the type I error rate for MOB and MOB-dS is shown in Figure \ref{fig:TypeIerror_rho05}.

\bibliography{mybib_els.bib}
\end{document}